\input tcilatex
\documentstyle[12pt]{article}
\begin{document}
\def\theequation{\arabic{section}.\arabic{equation}}

\title{ ASYMPTOTICS FOR SOLUTION TO THE CAUCHY PROBLEM FOR 
 VOLTERRA LATTICE WITH STEP-LIKE INITIAL VALUES} 

\author{V.L.Vereschagin} 

\date{\null} 

\maketitle \abstract{The connection between modulated Riemann surface 
of genus one and solution to Volterra lattice that tends to constants 
at infinity is studied. The main term of asymptotics for large 
time of solution to the mentioned Cauchy problem is written 
out.}

\vskip.5cm
\section{Introduction}
\setcounter{equation}{0}
\vskip.5cm

The Volterra lattice

\begin{equation}
\label{1.1}D_tc_n=c_n(c_{n+1}-c_{n-1}),\ c_n=c_n(t),\ n\in {\bf Z}
\end{equation}
is known as an interesting example of integrable difference- 
differential model with various applications to plasma physics, 
crystallography, zoology. The Inverse Scattering Data Method (ISDM) 
allows to investigate in details Cauchy problem for the Volterra lattice 
in case of quickly decreasing initial data (see \cite{man}) and in 
periodical case (\cite{ves}, \cite{ver1}). The further interest is 
stimulated by problems with more complex behavior of solutions to 
(\ref{1.1}) as $n\rightarrow \pm\infty$.

Examine the Cauchy problem for eq (\ref{1.1}) with initial data (i.d.) 
$c_n (0)$ which quickly tend to constant values as $n\rightarrow 
\pm\infty$:

\begin{equation} \label{1.2}\left\{
\begin{array}{c} u_n(0)=c_{2n}(0)=u^{\pm } \\
v_n(0)=c_{2n+1}(0)=v^{\pm }
\end{array}
,
\right \}
\ n\rightarrow \pm \infty \end{equation}
The goal of this paper is to construct asymptotics as $t\rightarrow
\infty$ for solution of problem (\ref{1.1})-(\ref{1.2}). Some 
similar problems were studied for Korteweg-de Vries equation (KdV)
(\cite{bik1}), Nonlinear Schrodinger equation (\cite{bik2}), 
Modified KdV (\cite{bik3}). The problem (\ref{1.1})-(\ref{1.2}) 
was posed in \cite{ver2} where the author formulated hypothesis to be 
proved lower.

The object of study of the next paragraph is a linear equation 
associated with Volterra lattice together with appropriate initial 
data for using the ISDM. In the third paragraph we formulate Riemann 
problem corresponding to initial conditions (\ref{1.2}) as $u^{\pm}= 
v^{\pm}= c^{\pm}$. The last restriction makes calculations much less 
tedious whereas all the results can be easily expanded to the general 
case $u^{\pm}\ne v^{\pm}$.

In the fourth paragraph we describe one-gap solutions to the Volterra 
lattice, modulated Riemann surface and its connection with the one-gap 
solutions.

The last paragraph contains asymptotic as $t\rightarrow \infty$ 
investigation of the formulated above Riemann problem. We prove a 
statement on representation of the main term of asymptotic solution to 
the Cauchy problem (\ref{1.1})-(\ref{1.2}) by modulated one-gap 
solution to the Volterra lattice.

\vskip.5cm
\section{Inverse Scattering Data Method} 
\setcounter{equation}{0}
\vskip.5cm

Examine a linear problem
\begin{equation}
\label{2.1}c_{n-1}^{1/2}\psi _{n-1}+c_n^{1/2}\psi _{n+1}=\lambda \psi _n,\
c_n\rightarrow c_{\pm },\ n\rightarrow \pm \infty ,\ c_{+}>c_{-}>0,
\end{equation}
$c_n$ are solutions to Volterra lattice, $\lambda$ is a spectral 
parameter. Define Riemann surface of genus 0 $\Gamma _{\pm }(z_{\pm 
},\lambda ):\ 2\lambda c_{\pm }^{-1/2}=z_{\pm }+z_{\pm }^{-1}.$ 
Involution $\sigma (z_{\pm })=z_{\pm }^{-1}$ near circle $\vert z 
\vert =1$ induces change of sheet of the surface $\Gamma _{\pm }$ 
near the cut $[0,c_{\pm}]=E_{\pm}$.

Fix solutions of problem (\ref{2.1}) by asymptotics

\begin{equation}
\label{2.2}\psi _n^{\pm }(\lambda )\sim e_n^{\pm }(z_{\pm })=z_{\pm }^{\pm
n}\exp \left( \frac 18tc_{\pm }\left( z_{\pm }^2-z_{\pm }^{-2}\right)
\right) ,\ n\rightarrow \pm \infty ,
\end{equation}
$\psi _n^{+}(P)$ is analytic on the lower sheet $\Gamma _{\pm }^- 
\setminus \infty ^-$ while $\psi _n^{-}(P)$ is analytic on the upper 
sheet $\Gamma _{\pm }^+ \setminus \infty ^+$. In the spectral domain 
$E_{+}$ there is a scattering correlation

\begin{equation}
\label{2.3}\psi _n^{+}(P)=a(P)\psi _n^{-}(P)+b(P)\psi _n^{-}(\sigma P),\
P\in E_{+}.
\end{equation}
Quantities $a(P),\ b(P)$ are connected by condition

\begin{equation}
\label{2.4}a(P)a(\sigma P)-b(P)b(\sigma P)=\frac{c_{-}^{1/2}W(\psi
_n^{+}(P),\psi _n^{+}(\sigma P))}{c_{+}^{1/2}W(\psi _n^{-}(P),\psi
_n^{-}(\sigma P))}=-\frac{z_{+}-z_{+}^{-1}}{z_{-}-z_{-}^{-1}},
\end{equation}
$W\left( \psi _n,\phi _n\right) =c_n^{1/2}(\psi _n\phi _{n+1-}\psi
_{n+1}\phi _n)$ is a Wronskian. There is also equality

\begin{equation}
\label{2.5}a(P)=\frac{W(\psi _n^{+}(P),\psi _n^{-}(\sigma P))}{
c_{-}^{1/2}(z_{-}-z_{-}^{-1})},\ \ b(P)=\frac{W(\psi _n^{+}(P),\psi
_n^{-}(P))}{c_{-}^{1/2}(z_{-}-z_{-}^{-1})},
\end{equation}
which signifies that function $a(P)$ can be analytically expanded to 
the lower sheet $\Gamma ^-$.

We suppose the spectrum to be solitonless, i.e. $a(P)$ has no zeroes 
on $\Gamma ^-$. One easily verifies that under $P \in E _{*}=E _{+} 
\setminus E _{-}$ the condition $a(P)=b(\sigma P)$ holds whence (as 
well as from (\ref{2.4})) one obtains:

\begin{equation}
\label{2.6}r(P)r(\sigma P)=1,\ P\in E_{*} ,
\end{equation}
where $r(P)=b(P)/a(P)$ is reflection coefficient.

\vskip.5cm
\section{Riemann problem.}
\setcounter{equation}{0}
\vskip.5cm

There is a general scheme for investigation of integrable systems via 
solving the inverse scattering problem in the form of matrix Riemann 
problem (MRP).

1. Let $\Psi_n (P)$ be 2$\ast$2-matrix piecewise analytic in $\bf C$ 
function with asymptotics under $P \rightarrow \infty _{\pm}$ 

\begin{equation}
\label{3.1}\Psi _n(P)=A_{\pm }\left( I+\lambda ^{-1}V_n^{\pm }+O(\lambda
^{-2})\right) \left(
\begin{array}{cc}
\left( \lambda c_{\pm }^{-1/2}\right) ^{\mp n} & 0 \\
0 & \left( \lambda c_{\pm }^{-1/2}\right) ^{\pm n}
\end{array}
\right) ,\ \lambda \rightarrow \infty _{\pm }
\end{equation}
where

$$
A_{\pm }=\left(
\begin{array}{cc}
0 & 1 \\
\frac 1{\sqrt{\frac{c_{\pm }}{c_{\mp }}}-1} & 0
\end{array}
\right) .
$$

2. Let function $\Psi _n (P)$ be analytic outside of the spectrum $E 
_{+}$ and on cut $E _{+}$ have a jump

\begin{equation}
\label{3.2}\Psi _n(P-i0)=\Psi _n(P+i0)G(P),\ P\in
E_{+}, where
\end{equation}

\begin{equation}
\label{3.3}G(P)=-\left(
\begin{array}{cc}
r^{-}(P) & \Delta ^{-} \\
\frac{r(P)r(\sigma P)-1}{\Delta ^{-}} & r^{-}(\sigma P)
\end{array}
\right) ,\ \Delta =c_{-}^{1/2}(z_{-}-z_{-}^{-1}),
\end{equation}
the upper index (-) signifies value on lower side of cut $E _{+}$. 
It follows from (\ref{2.6}) that matrix $G(P)$ becomes 
triangular in spectral branch $E _{*}$.

3. Non-degenerate matrix $\Psi _n (P)$ is a regular function in $\bf 
C$.

\underline{Statement 3.1.} Solution $c_n$ of Cauchy problem 
(\ref{1.1}) - (\ref{1.2}) for Volterra lattice is determined by 
solution $\Psi _n(P)$ of formulated above MRP (\ref{3.1})- (\ref{3.2}) 
via the following formula:

\begin{equation}
\label{3.4}c_n=\left[ \frac{\left( V_{n-1}^{+}\right) ^{21}}{\left(
V_n^{+}\right) ^{21}}\right] ^2,
\end{equation}
where $V^+$ is matrix coefficient of expansion (\ref{3.1}).

Proof. One must just verify that exact solution to MRP (\ref{3.1})
- (\ref{3.2}) is given by matrix

\begin{equation}
\label{3.5}\Psi _n(P)=\left(
\begin{array}{cc}
\psi _n^{+}(P) & \psi _n^{-}(\sigma P) \\
\psi _{n+1}^{+}(P) & \psi _{n+1}^{-}(\sigma P)
\end{array}
\right) \left(
\begin{array}{cc}
W^{-1}\left( \psi _n^{+}(P),\psi _n^{-}(\sigma P)\right) & 0 \\
0 & 1
\end{array}
\right)
\end{equation}

\underline{Statement 3.2.} Solution of the MRP (\ref{3.1}) - 
(\ref{3.2}) is unique.

Proof. Let $\Psi$ and $\widetilde{\Psi}$ be two solutions of MRP 
(\ref{3.1}) - (\ref{3.2}). According to Liouville theorem the matrix 
$\Psi (P) \widetilde{\Psi} ^{-1}(P)$ does not depend on $P$. Thus, the 
quantity $V _n ^{21}$ is derived from the MRP solution uniquely.

Investigation of solution to Volterra lattice as $t \rightarrow 
\infty$ implies asymptotic analysis of MRP (\ref{3.1}) - (\ref{3.2}).

\vskip.5cm
\section{One-gap solutions of Volterra lattice and modulation equations} 
\setcounter{equation}{0} 
\vskip.5cm

Exact formulas for real solutions to Volterra lattice were written out 
in paper \cite{ver1}:

\begin{equation}
\label{4.1}
\begin{array}{c} c_{2n}=u_n=u(\tau )=\zeta (2\omega \tau )-\zeta (2\omega
\tau -a_{-})-\zeta (a_{+})-\zeta (a_{-}-a_{+}), \\
c_{2n+1}=v_n=v(\tau )=\zeta (2\omega \tau -a_{+})-\zeta (2\omega \tau
)+\zeta (a_{-})-\zeta (a_{-}-a_{+}),
\end{array}
\end{equation}
where $\tau =\frac 12(a_{-}-a_{+})n+t+\omega ^{\prime };\ \zeta (x)$ 
is Weierstrass zeta-function:

$\zeta ^{\prime }(x)=-\wp (x),\ x=\int\nolimits_\infty ^{\wp (x)}\frac{d\nu
}{\sqrt{4\nu ^3-g_2\nu -g_3}};\ \wp (x+2\omega )=\wp (x);\ \wp (x+2\omega
^{\prime })=\wp (x);$
$a_{\pm },\ \omega ,\ i\omega ^{\prime }$ are four real parameters 
completely determining formulas (\ref{4.1}). Define the following four 
quantities:

\begin{equation}
\label{4.2}r_j=2\zeta \left( \frac 12(a_{-}-a_{+})+\omega _j\right) -2\eta
_j+\zeta (a_{+})-\zeta (a_{-}),\ j=1,2,3,4,
\end{equation}
where
$\omega _1=0,\ \omega _2=\omega ,\ \omega _3=\omega +\omega ^{\prime
},\ \omega _4=\omega ^{\prime },\ \ \eta _j=\zeta (\omega _j).$

The four parameters $\overline{r}=(r_1,r_2,r_3,r_4)$ also completely 
specifies solutions (\ref{4.1}) $u_{n}=u(\tau ,\overline{r}), \ 
v_{n}=v(\tau ,\overline{r}).$

Suppose quantities $\overline{r}$ depend on variables $x \in {\bf R}, 
t:\ \overline{r}=\overline{r}(x,t)$.

\underline{Definition.} On Riemann surface $\widetilde{\Gamma 
}(w,\lambda ):\ w^2=R_4(\lambda )=(\lambda -r_1)(\lambda -r_2)(\lambda 
-r_3)(\lambda -r_4)$ we define a couple of Abel differentials $\Omega 
_0,\ \Omega _1:$

\begin{equation}
\label{4.3}\Omega _0=\frac 12(\lambda +d_0)w^{-1}d\lambda ,\ \ \Omega
_1=\frac 12\left( \lambda ^2-\frac 12\Lambda \lambda +d_1\right)
w^{-1}d\lambda ,
\end{equation}
where $\Lambda =r_1+r_2+r_3+r_4;\ d_0,d_1$ are constants fixed by 
conditions

\begin{equation}
\label{4.4}\int\nolimits_{r_2}^{r_3}\Omega_0 
=\int\nolimits_{r_2}^{r_3}\Omega _1=0 \end{equation}

Equation

\begin{equation}
\label{4.5}D_t\Omega _0=D_x\Omega _1
\end{equation}
is called Whitham equation or modulation equation.

Formula (\ref{4.5}) can be rewritten in the form of system of 
quasilinear differential equations for which the quantities $r_j,\ 
j=1,2,3,4$ are Riemann invariants:

\begin{equation}
\label{4.6}D_tr_j=W_j(\overline{r})D_xr_j,\ j=1,2,3,4,
\end{equation}
where $W_j(\overline{r})=\Omega _1/\Omega _0(r_j).$ In paper 
\cite{ver2} characteristic velocities $W_j$ were written out 
explicitly in terms of complete elliptic integrals. In the same paper 
self-similar solutions to system (\ref{4.6}) were computed. Here we 
are interested in solutions that depend only on unique variable $\xi 
=x/t:\ \overline{r}=\overline{r}(\xi)$.

Define Riemann surface $\Gamma (w,\lambda ;\xi )$ in the following 
way:

\begin{equation} \label{4.7}\Gamma (w,\lambda ;\xi )=\left\{
\begin{array}{c}  w^2=(\lambda -r_1)(\lambda
-r_2^{-}),\ \xi \leq \xi ^{-}, \\
w^2=(\lambda -r_1)(\lambda -r_2^{-})(\lambda -r_3^{-}(\xi ))(\lambda
-r_4),\ \xi ^{-}<\xi \leq \xi _0^{-}, \\
w^2=(\lambda -r_1)(\lambda
-r_2^{-}(\xi ))^2(\lambda -r_4),\ \xi _0^{-}<\xi \leq \xi _0^{+}, \\
w^2=(\lambda -r_1)(\lambda -r_2^{+}(\xi ))(\lambda -r_3^{+})(\lambda
-r_4),\ \xi _0^{+}<\xi \leq \xi ^{+},
\\
\ w^2=(\lambda -r_3^{+})(\lambda
-r_4),\ \xi >\xi ^{+},
\end{array}\right.
\end{equation}
where $r_1\leq r_2^{+}(\xi )\leq r_3^{+}\leq r_2^0(\xi )\leq r_2^{-}\leq
r_3^{-}(\xi )\leq r_4;\ r_1,r_3^{+},r_2^{-},r_4$ are constants,

$$
\xi ^{-}=-\frac{\left[ (r_4-r_1)(r_4-r_2^{-})\right] ^{1/2}}{\log \left(
\frac{\sqrt{r_4-r_1}+\sqrt{r_4-r_2^{-}}}{\sqrt{r_4-r_1}-\sqrt{r_4-r_2^{-}}}
\right) },
$$

$$
\xi ^{+}=\frac 12(r_3^{+}+r_4-2r_1)-2\frac{(r_3^{+}-r_1)(r_4-r_1)}{
r_4+r_3^{+}-2r_1}+\frac{r_3^{+}-r_1}{\log \left( \frac 12\frac{r_4-r_3^{+}}{
r_4-r_1}\right) },
$$

$$
\xi _0^{-}=\frac 12(r_1+r_4)-r_2^{-},\ \ \xi _0^{+}=\frac
12(r_1+r_4)-r_3^{+}.
$$
Dependence of quantities $r_3^{-}(\xi ),r_2^0(\xi ),r_2^{+}(\xi )$ on 
variable $\xi$ is determined by self-similar solutions to system 
(\ref{4.5}):

\begin{equation}
\label{4.8}
\begin{array}{c}
\xi +W_3(r_1,r_2^{-},r_3^{-}(\xi ),r_4)=0,\ \xi ^{-}<\xi \leq \xi _0^{-}, \\
\xi +W_3(r_1,r_2^0(\xi ),r_2^0(\xi ),r_4)=0,\ \xi _0^{-}<\xi \leq 
\xi_0^{+} , \\ \xi +W_2(r_1,r_2^{+}(\xi ),r_3^{+},r_4)=0,\ \xi 
_0^{+}<\xi \leq \xi ^{+}.  
\end{array} 
\end{equation}

Quantities $\overline{r}=(r_1,r_2,r_3,r_4)$ as functions of variable 
$\xi$ are depicted in Fig.1.

%\begin{center}
%\begin{picture}(400,260)
%\put(90,230){\special{em:graph pic11.pcx} }
%\put(100,0){\parbox [c]{260pt}{{  \bf Fig.1.} }}
%\end{picture}
%\end{center}

The Riemann curve combined in such a way is a generalization of that 
obtained in paper \cite{ver2}. Define basic cycles as it is shown in 
Fig.2.

%\begin{center}
%\begin{picture}(400,260)
%\put(90,230){\special{em:graph pic12.pcx} }
%\put(100,0){\parbox [c]{260pt}{{  \bf Fig.2.} }}
%\end{picture}
%\end{center}

Holomorphic differential $\Omega = Dw^{-1} d\lambda$ ,$D$ is 
a constant fixed by condition $\oint\nolimits_a\Omega =1$.

Examine the following ansatz for solutions of Volterra lattice: 

\begin{equation}
\label{4.9}\left(
\begin{array}{c}
u(\tau ) \\
v(\tau )
\end{array}
\right) =\overline{c}(\tau \mid \Gamma (w,\lambda ; \xi 
))_{x=n}+O(t^{-\delta }),\ \delta >0, 
\end{equation}

$$
\overline{c}=\left\{
\begin{array}{c}
\left(
\begin{array}{c}
u \\
v
\end{array}
\right) ^{\pm },\ \xi >(<)\xi ^{\pm } \\
\left(
\begin{array}{c}
u(\tau ,
\overline{r}(\xi )) \\ v(\tau ,\overline{r}(\xi ))
\end{array}
\right) ,\ \xi ^{-}\leq \xi \leq \xi ^{+} ,
\end{array}\right.
$$
$\overline{r}(\xi )$ are branch points of curve $\Gamma (w,\lambda 
;\xi )$. Quantities $u^{\pm },v^{\pm }, \overline{r}$ are connected in 
the following way:

\begin{equation}
\label{4.10}
\begin{array}{c}
4r_1=u^{-}+v^{-}-2
\sqrt{u^{-}v^{-}},\ \ 4r_2^{-}=u^{-}+v^{-}+2\sqrt{u^{-}v^{-}}, \\
4r_3^{+}=u^{+}+v^{+}-2\sqrt{u^{+}v^{+}},\ \ 4r_4=u^{+}+v^{+}+2\sqrt{
u^{+}v^{+}},
\end{array}
\end{equation}
whereas one suppose that $r_1\leq r_3^{+}$.

Investigations in this and the next paragraphs deal with case 
$u^{\pm }=v^{\pm }=c^{\pm },\ r_1=r_3^{+}=0,\ a_{+}=-a_{-}$ which does 
not change principal scheme but considerably simplifies calculations. 

\vskip.5cm
\section{Asymptotic solving the Riemann problem.} 
\setcounter{equation}{0} 
\vskip.5cm

In paper \cite{ver3} explicit formulas for the case of one-gap 
spectrum of linear problem (\ref{2.1})were obtained. These formulas 
represent Baker-Akhiezer function in terms of elliptic functions:

\begin{equation}
\label{5.1}\widetilde{e}_n^{\pm }(t,z)=\gamma _n^{\pm 1}\exp \left[ \pm \pi
i\left( t{\cal V}(z)+n{\cal} P(z)\right) \right] \frac{\theta (\omega
^{\prime }\tau \pm z+P)}{\theta (\pm z+P)\theta (\omega ^{\prime }\tau -a+P)}
,
\end{equation}
where $a=a_{+}=-a_{-},\ P=\omega +a,\ \theta (\tau )=\sigma (\tau +\omega
)\exp \left( -\tau \eta -\frac{\tau ^2\eta ^{\prime }}{2\omega ^{\prime
}}\right) ,\ \tau =an+Vt,\ \frac{\sigma ^{\prime }(\tau )}{\sigma (\tau
)}=\zeta (\tau );$ variable $z$ is connected with parameter $\lambda$ 
via uniformization

$$
\lambda (z)=\zeta (z-a)-\zeta (z+a)+2\zeta (a)
$$

\begin{equation}
\label{5.2}{\cal V}(z)=\int\nolimits_{r_1}^{\lambda
(z)}\Omega _1,\ \ {\cal P}(z)=\int\nolimits_{r_1}^{\lambda (z)}\Omega _0,
\end{equation}

$$
\gamma _n=\beta ^n\left[ \frac{\sigma (\omega ^{\prime }\tau +\omega
)\sigma (2\omega ^{\prime }a+\omega )}{\sigma (\omega ^{\prime }(\tau
+2a)+\omega )\sigma (\omega )}\right] ^{1/2},\ \ \beta =\sigma (2\omega
^{\prime }a+\omega )/V
$$

Examine function

\begin{equation}
\label{5.3}\Phi _n(P)=\left\{
\begin{array}{c}
a^{-1}(P)\psi _n^{+}(P),\ P\in \Gamma ^{-} \\
\psi _n^{-}(P),\ P\in \Gamma ^{+} .
\end{array}
\right\}
\end{equation}
It is analytic on each sheet of the curve and has a jump on contour 
$\partial \Gamma ^{+}=E_{+}$ equal to $r(P)\psi _n^{-}(\sigma P)$. 
So the function $\Phi _n(P)$ can be restored by formula 

\begin{equation}
\label{5.4}\Phi _n(P)=e_n^{-}(P)+\frac 1{2\pi i}\int_{\partial \Gamma
^{+}}M(P,Q)r(Q)\psi _n^{-}(\sigma Q),
\end{equation}
$M(P,Q)$ is Cauchy kernel:

\begin{equation}
\label{5.5}M(P,Q)=-\frac{W(e_n^{-}(P),e_n^{-}(\sigma Q))}{
W(e_n^{-}(Q),e_n^{-}(\sigma Q))}\frac{d\lambda (Q)}{\lambda (P)-\lambda (Q)}
.
\end{equation}

Suppose now that $\xi \in \left[ \xi ^{-},\xi_0^{-} \right] $ and 
dynamics of the branch point $r^{-}_{3}(\xi)$ is unknown. Then 
function $\Phi _n(P)$ is specified by explicit formula via Cauchy 
integral on curve $\widetilde{\Gamma }$ (\ref{4.7}):

\begin{equation}
\label{5.6}\Phi _n(P)=e_n^{-}(P)+\frac 1{2\pi i}\int_{\partial \widetilde{
\Gamma ^{+}}\cup L}\widetilde{M}(P,Q)\widetilde{f}(Q),
\end{equation}
where
$\partial \widetilde{\Gamma }^{+}=\left[ r_1,r_2^{-}\right] \cup \left[
r_3^{-}(\xi ),r_4\right] ;\ \ L=\left[ r_2^{-},r_3^{-}(\xi )\right] \cup
\sigma \left( [r_3^{-}(\xi ),r_2^{-}]\right) ,\ \sigma$ is an 
involution that changes sheet on $\widetilde{\Gamma }$,

\begin{equation}
\label{5.7}\widetilde{M}(P,Q)=-\frac{W(\widetilde{e}_n^{-}(P),\widetilde{e}
_n^{-}(\sigma Q))}{W(\widetilde{e}_n^{-}(Q),\widetilde{e}_n^{-}(\sigma Q))}
\frac{d\lambda (Q)}{\lambda (P)-\lambda (Q)}
\end{equation}
Choice of function $\widetilde{f}$ is inspired by analogy with 
Zakharov-Manakov ansatz from paper \cite{zak-man}:

\begin{equation}
\label{5.8}\psi _n^{-}(P)=A(P)\widetilde{e}_n^{-}(P)+B(P)\widetilde{e}
_n^{-}(\sigma P) .
\end{equation}
Substitute this ansatz into the linear equation and get:

\begin{equation}
\label{5.9}
\begin{array}{c}
A(P)
\widetilde{e}_n^{-}(P)+B(P)\widetilde{e}_n^{-}(\sigma P)= \\
e_n^{-}(P)+\frac 1{2\pi i}\int\nolimits_{\partial \widetilde{\Gamma }
^{+}\cup L}\widetilde{M}(P,Q)r(Q)\left[ A(\sigma Q)\widetilde{e}
_n^{-}(\sigma Q)+B(\sigma Q)\widetilde{e}_n^{-}(Q)\right]
\end{array}
\end{equation}
Examine equation (\ref{5.9}) as $n \rightarrow \infty$ and use the 
known formula ($t \rightarrow \infty$)

$$
\frac 1{2\pi i}\int\nolimits_{-\infty }^\infty \frac{f(\eta )}{\eta -\mu +i0}
\exp (it\gamma (\eta ))d\eta =-f(\mu )\exp (it\gamma (\mu ))H(-\gamma
^{\prime }(\mu ))+O\left( \frac 1{\sqrt{t}(\mu -\mu _{CT})}\right) ,
$$
where $\gamma ^{\prime }(\mu _{CT})=0,\ H(x)=(1+signx)/2$ is Heaviside 
function. The role of $\gamma (\mu )$ is played by differential 
$\Omega _0(P)\xi -\Omega _1(P).$

Therefore one obtains:

\begin{equation}
\label{5.10}B(P)=-r(P)A(\sigma
P)H(\lambda (P)-\lambda (P_{CT})),\ P\in \partial \widetilde{\Gamma }^{+},
\end{equation}
where point $P_{CT}$ is determined by condition

\begin{equation}
\label{5.11}\Omega _0(P_{CT})\xi =\Omega _1(P_{CT}).
\end{equation}
Examine now (\ref{5.9}) as $n \rightarrow - \infty$ and get integral 
equations on $A(P)$:

\begin{equation} \label{5.12}
\begin{array}{c} A(P)
\widetilde{e}_n^{-}(P)=e_n^{-}(P)+\frac 1{2\pi i}\int\nolimits_{\partial
\widetilde{\Gamma }^{+}\cup L}\widetilde{M}(P,Q)r(Q)r(\sigma Q)A(Q)
\widetilde{e}_n^{-}(Q)H(\lambda (P)-\lambda (P_{CT})), \\ P\in \partial
\widetilde{\Gamma }^{+},
\end{array}
\end{equation}

\begin{equation}
\label{5.13}A(P)\widetilde{e}_n^{-}(P)=e_n^{-}(P)+\frac 1{2\pi
i}\int\nolimits_{\partial \widetilde{\Gamma }^{+}\cup L}\widetilde{M}
(P,\sigma Q)r(\sigma Q)A(Q)\widetilde{e}_n^{-}(Q),\ P\in L
\end{equation}
Thus the function $\widetilde{f}$ in (\ref{5.6}) must look in the 
following way:

\begin{equation}
\label{5.14}\widetilde{f}(P)=\left\{
\begin{array}{c}
-r(P)\left[ A(\sigma P)
\widetilde{e}_n^{-}(\sigma P)+B(\sigma P)\widetilde{e}_n^{-}(P)\right] ,\
P\in \partial \widetilde{\Gamma }^{+}, \\ (1+r^{+}(\sigma P))A(P)
\widetilde{e}_n^{-}(P),\ P\in \left[ r_2^{-},r_3^{-}(\xi )\right] , \\
(1-r^{+}(P))A(P)\widetilde{e}_n^{-}(P),\ P\in \sigma (\left[
r_2^{-},r_3^{-}(\xi )\right] ).
\end{array}\right.
\end{equation}
The quantity $A(P)$ should be searched from eqs (\ref{5.12}) - 
(\ref{5.13}). They indicate that function $A(P)$ is analytic on sheets 
$\Gamma _{\pm}$ and has a jump on contour $\partial \widetilde{\Gamma 
}^{+}\cup L$:

\begin{equation}
\label{5.15}A_n^{-}(P)=A_n^{+}(P)g(P),\ P\in \partial \widetilde{\Gamma }
^{+}\cup L,
\end{equation}
where conjugation matrix $g(P)$ has the form:

\begin{equation}
\label{5.16}g(P)=\left\{
\begin{array}{c}
1-r(P)r(\sigma P))H(\lambda (P)-\lambda (P_{CT})),\ P\in \partial
\widetilde{\Gamma }^{+}, \\ -r^{+}(\sigma P),\ P\in \left[
r_2^{-},r_3^{-}(\xi )\right] , \\
r^{+}(P),\ P\in \sigma (\left[ r_2^{-},r_3^{-}(\xi )\right] ).
\end{array}\right.
\end{equation}
Solution to such a problem is given by the following formula:

\begin{equation}
\label{5.17}A(P)=R(P)\exp \left( -\frac 1{2\pi i}\int\nolimits_{\partial
\widetilde{\Gamma }^{+}\cup L}\widehat{M}(P,Q)\log g(Q)\right) ,
\end{equation}
where $\widehat{M}(P,Q)$ is an analogue of Cauchy kernel:

\begin{equation}
\label{5.18}\widehat{M}(P,Q)=\int_{\infty _{+}}^Pm(P^{\prime },Q),
\end{equation}
$m(P,Q)$ is a meromorphic bidifferential fixed by conditions

\begin{equation}
\label{5.19}\oint_am(P,Q)=0,\ m(P,Q)\sim \frac{dq(P)dq(Q)}{(q(P)-q(Q))^2},\
P\sim Q,
\end{equation}
$q$ is a local parameter. It ensues from norm conditions (\ref{5.19}) 
that $\oint_bm(P,Q)=2\pi i\Omega (P)$.

Multiplier $R(P)$ is specified by condition of uniqueness of $A(P)$ 
and asymptotics $A(P)\rightarrow c_{-}^{1/2},\ P\rightarrow 
\infty _{\pm }:$

\begin{equation}
\label{5.20}R(P)=const\frac{\theta \left( \int_{\infty _{+}}^P\Omega
+D\right) }{\theta \left( \int_{\infty _{+}}^P\Omega +D_0\right) },
\end{equation}
where $D_0=-\left( \int_{\infty _{+}}^{r_1}+\int_{\infty _{+}}^{r_4}\right)
\Omega -K$, $K$ is a vector of Riemann constants (see \cite{ves}).

The conjugation conditions (\ref{5.16}) provide a formula for phase 
shift:

\begin{equation}
\label{5.21}D=D_0-\frac 1{2\pi i}\int\nolimits_{\partial \widetilde{\Gamma
} ^{+}\cup L}\Omega (Q)\log g(Q).
\end{equation}

So we constructed function $\Psi _n (P)$ (\ref{3.5}) which in case 
$r_3^{-}(\xi )=\lambda (P_{CT})$ asymptotically satisfy the 
conjugation condition (\ref{3.2}) and estimate (\ref{3.1}).

\underline{Statement 5.1.} Under $r_3^{-}(\xi )=\lambda (P_{CT})$ the 
following condition holds:

$$
\Psi _n(P+i0)G(P)\Psi _n(P-i0)=I+\delta (P,\xi ,t),
$$
where small parameter $\delta$ has the form

$$
\delta (P,\xi ,t)=\left\{
\begin{array}{c}
O\left( t^{-\epsilon }(P-P_{CT})\right) ,\ \left| P-P_{CT}\right| \geq
t^{-\epsilon }, \\
O(1),\ \left| P-P_{CT}\right| <t^{-\epsilon },\ \epsilon >0.
\end{array}\right.
$$
Analogously one studies the Riemann problem (\ref{3.2}) for all values 
of parameter $\xi$.

Now use formula (\ref{3.4}) and obtain

\underline{Statement 5.2.} The main term of asymptotics as 
$t\rightarrow \infty$ of solution to Cauchy problem (\ref{1.1}) - 
(\ref{1.2}) for Volterra lattice under condition of solitonless 
spectrum is described by Whitham-modulated one-gap solution  
$c(n,t\mid \Gamma (\xi ),D(\xi ))$ (\ref{4.1}), (\ref{4.2}), where 
Riemann curve $\Gamma (\xi)$ is determined by formulas (\ref{4.7}), 
phase shift $D(\xi)$ - by formula (\ref{5.21}). The remaining terms of 
the asymptotic series decrease as powers in $t$.

\end{document}